\documentclass[12pt]{article}
\usepackage{latexsym}
\usepackage{amsmath}
\usepackage{amsbsy}
\usepackage{amssymb}

\usepackage[utf8]{inputenc}

\usepackage{verbatim}
\usepackage{bm}

\usepackage{curves}	
\usepackage{epic}
\usepackage{eepic}
\usepackage{epsfig}
\usepackage{wasysym}
\usepackage{pifont}

\usepackage{slashed}

\usepackage{tikz}
\usetikzlibrary{mindmap}
\usepackage{metalogo}

\usepackage{tikz}
\usetikzlibrary{mindmap}
\usepackage{capt-of}

\UseRawInputEncoding

\newcommand{\be}{\begin{equation}}
\newcommand{\bear}{\begin{eqnarray}}
\newcommand{\ear}{\end{eqnarray}}
\newcommand{\ee}{\end{equation}}

\newcommand{\bi}{\bibitem}

\newcommand{\vs}{\vspace}
\newcommand{\hs}{\hspace}

\usepackage{hyperref}

\usepackage{tcolorbox}

\newtcolorbox{mybox}{colback=red!5!white,
colframe=red!75!black}

\newtcbox{\mybox2}[1][]{colback=red!5!white,
colframe=red!75!black,fonttitle=\bfseries,
title=#1}

\begin{document}

\begin{center}

\

\vs{.8cm}

\baselineskip .7cm

{\bf \Large Extended Relativity: Beyond\footnote{This paper is the invited talk, reformated, given by the author, at IARD 2016.}} 

\vs{4mm}

\baselineskip .5cm
Juan Francisco Gonz\'alez Hern\'andez\footnote{e-mail: jfgh.teorfizikisto@gmail.com}\footnote{e-mail: juanfrancisco.gonzalez1@educa.madrid.org}\footnote{Department of Physics and Chemistry, IES Humanejos, Parla (Spain)}

\vs{3mm}

\end{center}
\abstract{Despite their success, General Relativity (GR) and the Standard Model (SM) are currently understood as effective field theories only valid up to some energy (length) scale, where new physics is expected to appear. We review the framework of extended relativity (ER) in Clifford spaces (C-spaces), summarizing some of its concepts, methods and results. We also discuss likely links of this approach with other relativities (OR), beyond GR and SM theories, recent ideas from emergent spacetime (ES) and quantum entanglement (QE), in search for quantum gravity (QG) and unification (U). Finally, we explore and expose the need to go even beyond ER, and the challenges it poses, from both experimental and theoretical sides. }
\baselineskip .43cm
{\footnotesize
}

\vs{3mm}

\hs{7mm}

\baselineskip .55cm

\section{Introduction and motivation}

Theoretical Physics at the beginning of the 21st century is coded into 2 big Effective Field Theories (EFT), at least up to energies tested in the colliders or the lab about $E\sim 100GeV-1TeV$,  and distances\footnote{Excepting gravity at small distances, hard to handle with.} in the range $10^ {-18}m<\lambda< R_{Obs}$ :
\begin{itemize}
\item \textbf{General Relativity} (GR) or Standard theory of Gravity (SG). This theory can be condensed in a set of pure field theory equations derived from a lagrangian density, and it describes gravity as curvature in the space-time geometry caused by fields of matter (energy-momentum) living on the space-time manifold,  as follows
\be \boxed{G_{\mu\nu}=\dfrac{8\pi G}{c^4}T_{\mu\nu}$ with $\mathcal{L}_{GR}=\mathcal{L}_{EH}+\mathcal{L}_M}
\ee 
\item \textbf{Quantum Field theory} (QFT), a gauge field theory based on the gauge group $G=SU(3)\times SU(2)\times U(1)$, and describing every fundamental force excepting gravity in a quantum way. This is called the \textit{Standard Model} (SM) or sometimes the \textit{Standard Theory} (ST). Despite its complexity, it can more or less be described by a lagrangian field theory as well, with some pieces and terms more mysterious and less tested than others:
\end{itemize}

\bear
 \boxed{L_{SM}=L_\psi+L_{gauge}+L_{Y}+L_{Higgs}}\\
 L_\psi=i\overline{\Psi}\slashed D\Psi +h.c.\
 L_{gauge}=-\dfrac{1}{4}F_{\mu\nu}F^{\mu\nu}\\
  L_{Y}=Y_{ij}\phi \overline{\Psi}_i\Psi_j+h.c.\\
  L_{Higgs}=\vert D_\mu\phi\vert^2-V(\phi)
\ear

\vs{4mm}

The impressive experimental success of these two big pictures and theories can only be confronted by their apparent incompatibility. To be more precise, more than incompatibility, some experimental and theoretical issues hint that they can not be the end of the history but only the current edge and limit of our physical knowledge. Current frontiers in our knowledge are triggered by certain (not complete) list of questions we can not answer in the framework of GR and the SM:

\begin{enumerate}
\item \textit{Dark Matter (DM) and Dark Energy(DE)}. What are they? Particles and/or Modified Gravity?
\item \textit{The Cosmological Constant} (dark energy?) problem.
\item \textit{The origin of mass} (how to explain why the Yukawa couplings and masses are those we observe?).
\item \textit{Black Hole (BH) physics}: where does BH entropy come from? The Black Hole information paradox problem and the fate of space-time singularities, \dots
\item \textit{Quantum Mechanics and its foundations}. Is it geometry? Is it (really) fundamental or emergent?
\item \textit{The emergence of space-time}. Is space-time itself fundamental or emergent?
\item \textit{Gravitational, strong and electroweak forces} \textbf{not unified}. What is the right Grand Unified Theory (GUT)? What is the right super-GUT theory or the theory of everything (TOE)?
\end{enumerate}

\vs{4mm} 

To address part of some of these problems, standard theories are NOT enough. We need to go \textit{beyond}. Thus, quantum gravity (QG) and unification (U), implies to build the so-called Beyond General Relativity (BGR) and Beyond Standard Model (BSM) theories. There are some popular candidates and well-known approaches to BGR and BSM. Let us write a brief selection of these ideas and models:

\begin{enumerate}
\item \textit{Superstrings/M-theory}.
\item \textit{Loop Quantum Gravity} (LQG).
\item \textit{CFT, GUT's, NC geometry, twistors}. 
\item \textit{Phenomenology of QG}.
\item \textit{Higher Spin Theories, Generalized UP, analogue models, SME}.
\item\textit{ World crystals, holography, gauge/gravity correspondence, emergent spacetime, QIT (Quantum Information Theory) or ``spacetime from entanglement'', \ldots}
\item Deformations and extensions of special relativity (SR), GR and Quantum Mechanics (QM). There are nontrivial extensions of relativity and other relativities (e.g., doubly/triply special relativity) out there and no one seems to pay attention to some of their ideas.
\end{enumerate}

In this article\footnote{It is a lightly enlarged version of a talk given by the author at IARD2016, Ljubljana (Slovenia), on June 6th, 2016.}, we approach mainly the path to unification of the so-called Extended Relativity (ER) in Clifford spaces (C-spaces), its ``state of art'' and its own ``beyond'', what could be called beyond Extended Relativity (BER).

\section{Bits on Extended Relativity}

Why to study a new approach? Why to learn Extended Relativity? Physics and Science are about ideas, models and theories passing experimental tests. There is nothing wrong to test new paths. Moreover, we can provide six good reasons to study Extended Relativity and Extended Relativity in C-spaces: 

\begin{itemize}
\item Not too many people out there doing it! No competitors! No matter if you don't like a model or not, not having too many competitors is a good reason. 
\item New ways to enlarge relativity/gauge symmetries. We suspect unification requires a higher notion of symmetry and maybe a new relativity principle. Indeed, string theory/M-theory and other main approaches still lack a unifying principle (cf. equivalence principle, Lorentz invariance, diffemorphism  invariance,\ldots)
\item Derive relationships, equivalences and dualities with other known major/minor approaches.
\item An alternative tool to compactification, extra dimensiones (ED), and U.  
\item Clifford algebras seem to be important too in QIT.
\item Create new predictions to be tested in experiments and explore new paths towards U. With the rise of neutrino and gravitational wave astronomy, and a new generation of experimental devices, we can reach the goal of testing the boundaries of our current theories, in order to find out how to supersede them all.  
\end{itemize}

\subsection{Polyvectors and the rise of ER ``machines''}

From \cite{Castro1, Castro2}, the work \cite{PavsicCliff},\cite{pezzaglia99}, we have a key principle: ER theory in C-spaces  generalizes of the notion of the interval in Minkowski space to a manifold we call Clifford space ($C$-space) and naturally requires extended objects.\footnote{\textit{Matej Pav\v{s}i\v c (IARD 2002)} introduced the idea of working in C-space: polydimensional relativity and C-space as the ``arena'' of physics.}\\

ER requires polyvectors (polyforms). What is a polyvector? The Clifford valued polyvector $X=X^ME_M$ is defined as:
\be X= X^ME_M=\sigma {\underline 1} + x^\mu \gamma_\mu +
x^{\mu\nu} \gamma_\mu \wedge \gamma_\nu +...+ x^{\mu_1 \mu_2
....\mu_D}
\gamma_{ \mu_1 } \wedge \gamma_{ \mu_2}....
\wedge \gamma_{ \mu_D }
\ee

\vs{2mm}

\textbf{ Interpretation:} a ``point'' in
$C$-space has coordinates $ X^M $ and basis $E_M$. The series ends  at a {\it finite}
grade depending on the dimension $D$. A Clifford algebra $ Cl (r, q) $
with $ r + q= D $ has $ 2^D $ basis elements. Clifford algebra (CA) or geometric calculus (GC) use the product $ab=a\cdot b+a\wedge b$. 

\vs{2mm}

ER implies the transition from Minkowski to Clifford space-time 
\begin{itemize}
\item For simplicity, the gammas
$\gamma^\mu$ correspond to a Clifford algebra associated with a
flat spacetime $ \lbrace \gamma^\mu, \gamma^\nu \rbrace = 2\eta^{\mu\nu} $. But we can use the construction with curved spacetimes as well. Introduce a
metric
 \[ \lbrace \gamma^\mu, \gamma^\nu \rbrace = 2g^{\mu\nu} \]
Einstein introduced the speed of light as a universal \textit{absolute} invariant in
order \textbf{to unite} space with time (to match units) in the
Minkwoski space interval:
$$ d s^2 = c^2 d t^2 + d x_i d x^i $$
Mimicking Einstein, the C-space interval merges objects with different dimensions, through the so-called C-space metric, a multi-index version of space-time. Therefore, the C-space interval generalizes Minkovskian spacetime:
  \[ dX^2 = d\sigma^2 + dx_\mu dx^\mu + d x_{\mu\nu} dx^{\mu\nu} +... \]
  
   \end{itemize}

We can see this with an alternative procedure:

\begin{enumerate}
\item Take the differential $ d X$ of $X$.  Compute the
scalar  $ < d X^{\dagger} dX >_0 \equiv
d X^{\dagger} * d X \equiv |d X|^2$ and obtain
the C-space extension of the particles proper time in Minkwoski space
\item 
The symbol $ X^{\dagger} $ denotes the {\it reversion} operation and involves
reversing the order of all the basis $ \gamma^\mu$ elements in the expansion of
$ X$.It is the analog of the transpose (Hermitian) conjugation
\item The C-space
metric associated with a polyparticle motion is :
\begin{equation}
|d X|^2 = G_{MN} \, d X^M d X^N 
\end{equation}
where $G_{MN} = E_M^{\dagger} * E_N$ is the $C$-space metric.
  \begin{equation}
|d X|^2= d \sigma^2 + L^{-2 } dx_\mu dx^\mu +
L^{-4 } d x_{\mu\nu} dx^{\mu\nu } +...+  L^{-2D} d x_{\mu_1...\mu_D}
\, d x^{\mu_1...\mu_D}
\end{equation}
\end{enumerate}

We need a new scale to match units. It seems to be something with dimensions of ``length''. Planck length is the natural choice. Thus:

\begin{itemize}
\item \textit{Neccesary} introduction:  \textbf{Planck scale $L$}.  It is
\textbf{length} parameter is needed in order to tie objects of different
dimensionality together: 0-loops, 1-loops,..., $p$-loops.
\item This procedure can be carried to all closed p-branes ($p$-loops)
where the
values of $p$ are $ p = 0, 1, 2, 3,... $. The $ p = 0 $ value
represents the
center of mass and the coordinates $ x^{\mu\nu}, x^{\mu\nu\rho}...$
\end{itemize}


\subsection{Motion in C-space}

We can study motion in C-space as in usual space-time, plus extras degrees of freedom. The line element and polymomentum read

\bear d X^A d X_A = d \sigma^2 + \left ({{d x^0}\over L} \right)^2
 - \left ({{d x^1}\over L} \right)^2   - 
     \left ({{ d x^{01}}\over L^2} \right)^2   ....  +
 \left ({{ d x^{12}}\over L^2} \right)^2 -
 \\
\left ({{d x^{123}}\over L^3 } \right)^2
 - \left ({{d x^{0123}}\over L^4} \right)^2 +... = 0  
 \ear

\begin{itemize}
\item Vanishing of ${\dot X}^B {\dot X}_B$ is equivalent to vanishing of the above
$C$-space line element and by ``..." we mean the terms with the remaining components such
as $x^2$, $x^{01}$, $x^{23}$,..., $x^{012}$, etc.
 \item The C-space metric is $G_{MN} =E_M^{\dagger} * E_N$ and if the dimension of spacetime is 4, then $x^{0123}$ is
the highest grade coordinate.
\end{itemize}

We can study the polyvelocity in C-space time:

\begin{eqnarray*}
V^2 = - \left (L {{d \sigma}\over {d t}} \right)^2 + \left (
{{d x^1}
 \over {d t}} \right)^2       +   \left ({{ d x^{01}}\over L^2} \right)^2       ...\\  - \left ({1\over L} {{d x^{12}}\over {d
t}}
 \right)^2 + \left ({1\over L^2} {{d x^{123}}\over {d t}} \right
)^2
 + \left ({1\over L^3} {{d x^{0123}}\over {d t}} \right)^2 -...
\end{eqnarray*}

 We find that
 
 \begin{itemize}
 \item The maximum speed  $ V^2 = c^2 $ in C-space depends on extra r-vector quantities. 
\item The maximum speed squared $V^2$ contains components of the 1-vector velocity $d x^1/ d t$,  but also the multivector $d x^{12}/ d t$, $ d x^{123}/ d t$, \ldots
The following special cases  in $C$-space are different from zero, are of particular interest:
\end{itemize}


Indeed, motion in C-space introduces some natural extra maxima, beyond the speed of light.

\begin{itemize} 
\item \textbf{Maximum 1-vector speed.}
 \[ {{d x^1}\over {d t}} = c = 3.0 \times 10^8 m/s  \]
 \item \textbf{ Maximum 3-vector speed.}
\[ {{d x^{123}}\over {d t}} = L^2 c = 7.7 \times 10^{-62} m^3/s  \]
\item \textbf{ Maximum 3-vector \textit{diameter} speed and Maximum 4-vector speed}
 \[ {{d \root 3 \of {x^{123}} }\over {d t}} = 4.3 \times 10^{-21}m/s  \;\;\;\;{{d x^{0123}}\over {d t}} = L^3 c = 1.2 \times 10^{-96}
  m^4/s 
 \]
\end{itemize}

\textbf{Remark:} it has not been noted before, to our knowledge,  that  you can also get maximal limits to $n$ order derivatives given by 
\[Max \left(\dfrac{d^{n+1}x}{dt^{n+1}}\right)\leq c\left(\dfrac{c}{L}\right)^n\]
Indeed, by duality, if valid as fundamental symmetry, minimal limits should be also considered. 
\[Min \left(\dfrac{d^{n+1}x}{dt^{n+1}}\right)\geq C\left(\dfrac{C}{l}\right)^n\]

\textbf{Remark (II)}: hint of a high derivative extension of relativity? Have you ever heard about tachyons and epitachyons?

\subsection{C-space Maxwell Electrodynamics}

We can study a C-space gauge field theory version of electromagnetism. Reasons:

\begin{enumerate}
\item C-space electrodynamics generalize Maxwell's theory: \[ F=dA, \; dF=0 \]
\item Abelian C-space electrodynamics is based on the polyvector field 
\be A=A_NE^N= \phi {\underline 1} + A_\mu \gamma^\mu + A_{\mu \nu} \gamma^\mu
\wedge \gamma^\nu +...=(\phi,A_{\mu}, A_{\mu \nu},...) 
\ee
\item Defining the C-space operator ($ M,N=1,2,\ldots,2^D $)
\[ d = E^M \partial_M = {\underline 1} \partial_\sigma + \gamma^\mu
\partial_{x_{\mu} } + \gamma^\mu \wedge \gamma^\nu \partial_{
x_{\mu\nu} } +... \]
\item The generalized field strength in C-space is:
\[  F = dA = E^M \partial_M (E^N A_N) = E^M E^N \partial_M
A_N =
 \]
 \[ \dfrac{1}{2}\left\lbrace E^M,E^N \right\rbrace \partial_MA_N +\dfrac{1}{2}\left[ E^M,E^N \right]\partial_MA_N= \]
 \[\dfrac{1}{2}F_{(MN)}\left\lbrace E^M,E^N \right\rbrace +\dfrac{1}{2}F_{[MN]}\left[ E^M,E^N \right]  \]

\end{enumerate}

%
C-space Maxwell Electrodynamics (CME) uses a decomposition in symmetric and antisymmetric parts of the strength field in C-space with the aid of geometric product

\[  F_{ (MN) } = { 1 \over 2 } (\partial_M A_N + \partial_N A_M) \;\; F_{ [MN] } = { 1 \over 2 } (\partial_M A_N - \partial_N A_M)
 \]
 and also has a C-space Maxwell-like action

 \[  I[A] = \int [D X] F_{ [MN]}F^{ [MN] }\]
 
  with measure  
  
   \[ [D X] \equiv (d \sigma)(d x^0 d x^1...) (d x^{01}d x^{02}...)....
(d x^{012...D }) \]

CME action is invariant under C-space gauge transformations

 \be  A'_M = A_M + \partial_M \Lambda \ee
 and the minimal matter-field coupling interacting term after absorbing constants is similar to the coupling of p-branes to antisymmetric fields, as those arising in superstrings or M-theory: 
 \[ \int A_M dX^M = \int [D X] J_M A^M \]

Naturally, we can build up CME actions, field equations and generalizations with non-abelian fields.

 \be  \partial_M F^{ [MN] } = J^N \;\; \partial_N \partial_M F^{ [MN] } = 0 = \partial_N J^N = 0 \ee

 In fact, the C-space Maxwell action is only a piece of the more general C-space action
 
 \be I[A] = \int [D X] \, F^{\dagger}*F =  \int [ { \cal D } X  ] ~ <F^{\dagger} F>_{scalar} \ee
 and the non abelian equations should be written as 
 
 \be F = DA = (d A + A \bullet A) \;\; E_M \bullet E_N = E_M E_N - (-1)^{ s_M s_N} E_N E_M \ee


The \text{C-space gauge fields} in general can be written as follows:
 \[{\bf X } ~ = ~ \varphi ~{\bf 1} ~ + ~ x_\mu ~\gamma^\mu ~ + ~ x_{\mu_1 \mu_2} ~\gamma^{\mu_1} \wedge \gamma^{ \mu_2 } ~ + ~ x_{\mu_1 \mu_2 \mu_3 } ~
 \gamma^{\mu_1} \wedge \gamma^{ \mu_2 } \wedge \gamma^{ \mu_3}  +   ... \]

\textit{Example:}  Polyvector valued gauge field in  $Cl (5,C)$  acquires the form
 $ {\cal A}_M ( { \bf X}  ) =  A_M^I ( { \bf X} ) ~\Gamma_I $  and is  
 spanned by $16 + 16$  generators. The expansion of the poly-vector $ {\cal A}_M^I $ is also of the form 
 $$ {\cal A}_M^I ~ = ~ \Phi^I  ~{\bf 1} ~ + ~ A_\mu^I ~\gamma^\mu ~ + ~ A_{\mu_1 \mu_2}^I 
 ~\gamma^{\mu_1} \wedge \gamma^{ \mu_2 } ~ + ~ A_{\mu_1 \mu_2 \mu_3 }^I ~
 \gamma^{\mu_1} \wedge \gamma^{ \mu_2 } \wedge \gamma^{ \mu_3} ~ + ~  ... $$

 In order to match units, a \textit{length scale} needs (again) to be introduced in the expasion. The Clifford-algebra-valued gauge field $ {\cal A}_\mu^I ( x^\mu) \Gamma_I $ in ordinary spacetime  is naturally embedded into a far richer object   $ {\cal A}_M^I ( { \bf X}  ) $ . The 
scalar $ \Phi^I $  admits the $ 2^5 = 32$ components 
$ \phi, ~ \phi^i, ~ \phi^{ [ ij ] }, \phi^{[  i jk  ] }, \phi^{[  i jk l ] }, \phi^{[  i j k l m ] } $  of  
$Cl ( 5, C ) $ space. 

\textit{Example (II)}: Field theory equations on C-space. C-space Klein-Gordon and Dirac Wave Equations can be derived from a sort of Polymomentum correspondence principle (POP)

\[ P_A \rightarrow - i {\partial \over {\partial X^A}} =
 -i \left ({\partial \over {\partial \sigma}}, {\partial \over {\partial x^\mu}},
 {\partial \over \partial x^{\mu \nu}},... \right) \quad \Psi(x^{\mu}) \rightarrow  \Psi(x^{A}) \]

The C-space Klein-Gordon wave equation reads
\be\left ({ \partial^2\over \partial \sigma^2 } +
{ \partial^2 \over \partial x^\mu \partial x_\mu } +
{ \partial^2 \over \partial x^{\mu \nu} \partial x_{\mu \nu} } +
... + M^2 \right) \Phi = 0 \ee

and C-space Dirac wave equation is 
\be - i \left ( {\partial \over \partial \sigma}  +
\gamma^\mu { \partial \over \partial x_\mu } +
\gamma^\mu \wedge \gamma^\nu { \partial \over \partial x_{\mu \nu} } +
... \right) \Psi = M \Psi \ee

Note we used natural units in which $\hbar = 1, c = 1$.


\section{\scshape Beyond ER: a ``new'' path to QG and U}
\subsection{Hints and Links}

The 1st link pointing out a new path is an old friend. The \textit{maximal acceleration} (tension, power, \ldots) principle. 
\begin{itemize}
\item Caianiello's QM \cite{Caianiello1981}as induced geometry in phase-spacetime includes
\[a_C=2\dfrac{mc^3}{\hbar}=2\dfrac{Ec}{\hbar}\]
\item Born's reciprocal relativity\cite{Born1,Born2}. 
\item Maximal Force implies a maximal power (e.g. Schiller 2006). The recent LIGO detection of GW is only about $10^{-2}$ the maximal power. \textbf{Hope:} we aspire to test maximal power (force?) with GW radiation in the future!
\item Maximal acceleration\cite{maxacc1,maxacc2}, via EP, implies a maximal, critical, strong gravitational field and force\cite{maxforce1, maxforce2, maxforce3, maxforce4, maxforce5}. The SM in strong fields (Schwinger effect\cite{SchwingerRev}) remains as an experimental challenge yet. 
\end{itemize}

The 2nd link is \textit{emergence}. Emergent spacetime and complexity is being more and more important since the appearance of the entanglement-geometry connection. 
\begin{itemize}
\item  Role of ``emergence'': emergent spacetime from entanglement? \textit{Is quantum entanglement the key?}
\item Complexity and gravity interplays. Indeed, Susskind et al. \cite{suss} recently related complexity with action and they got the rate:
\[\dfrac{d\mbox{Complexity}}{dt}\leq \dfrac{2M}{\pi \hbar}\]
It suggests a link with maximal acceleration as well after rescaling with $\pi, c^3$.
\end{itemize}

The 3rd link is mostly unknown. Perhaps, forgotten. There are other Relativities (OR) going beyond SR and GR that have been postulated during the 20th and 21st centuries. Past works on (mostly forgotten or uncommon) OR should be a topic for further research\footnote{Even the author has a little project about it, unpublished.}. A simple (non-exhaustive) list includes (choose one or many, as you wish):
\begin{itemize}
\item Born seminal work on reciprocal relativity\cite{Born1},\cite{Born2}.
\item Fantappie's final relativity \cite{fantappie54} and Arcidiacono's projective relativity\cite{Arcidiacono88} (dS like).
\item Kalitzin's multitemporal relativity\cite{kalitzin1} (see also the 1975 book \cite{kalitzin2}). 
\item Barashenkov's 6D relavitivy\cite{Barashenkov1,Barashenkov2,Barashenkov3}.
\item Cole's 6D spacetime relativity and cellular spacetime\cite{Cole1,Cole2,Cole3, Cole4}.
\item Bogoslovski's anisotropic relativity (Very Special Relativity).\cite{bogo1},\cite{bogo2},\cite{bogoASRbook}
\item De Sitter relativity \cite{dSrel1, dSrel2} (doubly and even triply SR has been discussed \cite{dSrel3}).
\item C. Nassif's minimal velocity relativity \cite{nassif1},\cite{nassif2}.
\item Gogberashvili's octonionic relativity\cite{gog1,gog2,gog3,gog4}.
\item A.A. Ketsaris 7d and 10d extended relativities\cite{ketsaris99a,ketsaris99b}.
\item Wilczek's total relativity\cite{wilTotal04}.
\item \ldots 
\end{itemize}
Are all wrong or some of their ideas could be right indeed some of them? Until now, as F. J. Dyson remarked\cite{dyson72}, they are forgotten lost opportunities. Forever?


\subsection{Beyond ER: issues and questions}

{Beyond ER: Hints of a new ER}
Everything so far sounds good, what is the problem with ER? A critical view:
  \begin{itemize}
  \item No clear principle(s) but points into it(them) in a sense: why is fundamental scale $L=L_p$? What about a dual extension with MAXIMAL/dS length $L=L_\Lambda$? 
  \item Transitions between different signatures not understood yet(even worst, no observational signature seen, and we observe a 3+1 universe, if we neglect that dark matter and dark energy puzzles).
  \item The Clifford group choice: we can not choose a reason of why to pick one instead any other.
  \item Similar issues to theories of strings/branes: no hints of the emergence of multiple energy or mass scales.
  \end{itemize}
  
 However, ER (or BER) gives hints and extra suggestions of how to proceed. We are not claiming ER are the final theory. We are only saying that some ideas could help us in the path towards it. 

Extra hints of a new ER are provided by Classical Mechanics and its generalizations. Classical Mechanics is based on the Poincare-Cartan two-form 
\be
\omega_2=dx\wedge dp
\ee

There, $p=\dot{x}$. Quantum Mechanics is secretly a subtle modification of this. By the other hand, the so-called Born-reciprocal relativity is based on the ``phase-space''-like metric
\[ds^2=dx^2-c^2dt^2+Adp^2-BdE^2\]
and its full space-time+phase-space extension:
\[ds^2=dX^2+dP^2=dx^\mu dx_\mu+\dfrac{1}{\lambda^2}dp^\nu dp_\nu\]
 The extension of Born's reciprocal relativity in C-spaces based on higher accelerations \textit{is} an interesting open problem. E.g.: take $ds^2=dx^2+dp^2+df^2$. We have an invariant and likely hidden Nambu dynamics
\be\omega_3=d X\wedge  dP\wedge dF\ee

\textbf{Question(not totally solved):} What is the symmetry group or invariance of the above $(n+1)$-form and whose intersection with the $SO(D(n+1)) $ group gives the higher order metaplectic group?
$$ \omega_{n+1}=dx\wedge dp\wedge d\dot{p}\wedge\cdots \wedge dp^{(n-1)}$$
where we include up to $(n-1)$ derivatives  or equivalently
$$ \omega_{n+1}=dx\wedge d\dot{x}\wedge d\ddot{x}\wedge\cdots \wedge dx^{(n)}$$

Is this argument asking for some kind of unified framework for higher derivative theories? It is natural to consider higher derivative corrections to the EFT of gravity (GR) and the SM. There has been renewed interest in this field the last years. Morevoer, higher derivative (even nonlocal infinite order theories) are thought to cure singularities, and maybe explain the mysteries of dark matter and dark energy as well.

\subsection{The UR conjecture}
``New'' relativities and some extensions of relativity do exist and they include several ingredients and hypothesis to be tested. Furthermore, I also propose that behind the BER is a new version of relativity and/or quantum field theory, trascending them all:

\begin{tcolorbox}
\begin{center}
\textbf{Ultimate Relativity conjecture}
\end{center}
 There is a beyond extended relativity (BER) principle with \textit{min/max} values of any \textit{n}-th derivative of C-space coordinates (also for any polyvector derivatives in C-spaces). The existence of this \textit{ultimate relativity} \textbf{(UR)} principle is linked to the nature of the maximal and minimal symmetry of the degrees of freedom of the theory and the own Nature and ``reality''. It claims a sort of polydimensional relativity and a polyvectorial relational Universe. 
\end{tcolorbox}

\textbf{UR:} german word, ``original''. Also, in Geology, ``the first (prime) supercontinent''.

\textbf{Remark:} This is not really ``completely new'' but a reboot and revival of an older idea, cf. ``final relativity'' (Fantappie, Arcidiacono) and more recently Wilczek's ``total relativity''.

\textbf{Remark (II)}: BER implies UR. 

Indeed, we could envision DM as minimal acceleration dynamics.  Suppose there is a minimal acceleration $a_0$ (minimal force $F_0$). Then: 
\be\dfrac{v^2}{R}=G\dfrac{M}{R^2}+a_0\ee
and from this, by simple squaring, you obtain
\be v^4=G^2M^2R^{-2}+a_0^2R^2+2GMa_0\ee
In the limit $G^2,a_0^2<<1$, you get the phenomenological law
\be\boxed{v^4=2GMa_0}\ee

\textbf{Idea:} DM, even if real, could be hinting a minimal acceleration (MOND-like) dynamics.

This argument can be generalized to the presence of dark energy. DM plus DE could be minimal acceleration dynamics plus maximal length. Suppose (with $c=1$) there is a minimal acceleration $a_0$ (minimal force $F_0$) and a cosmological constant (de Sitter radius) giving some class of maximal length. Then:
\be\dfrac{v^2}{R}=G\dfrac{M}{R^2}+a_0+\Lambda R\ee
and from this, by simple squaring, you obtain
\be v^4=G^2M^2R^{-2}+a_0^2R^2+2GMa_0+2GMR\Lambda+2a_0\Lambda R^3+\Lambda^2R^4\ee
In the limit $G^2, a_0^2, \Lambda<<1$, you get the phenomenological law
\be \boxed{v^4=2GMa_0}\ee

\textbf{Idea:} DM, DE, even if real, could be hinting a (MOND-like) minimal acceleration and a maximal length dynamics. Does MOND fail (when it does) because we ignore extra terms?


\section{\scshape Conclusions}

Where do we stand with ER and BER? At current time, no experimental hints of them so far. However, we could stress some points to consider the study of these theories:

\begin{enumerate}
\item There are multiple advantages of recurring to $C$-spaces. Not covered here: gravity with torsion, YM fields and nonabelian EM with CS terms,\ldots
\item Every physical quantity is a polyvector. ER implies the rise of ``polyvector/polyform'' machines. Polydimensional and signature relativity should be included as a part of the theory. Dualities between theories of different dimensionalities have a functorial origin.
\item C-space dynamics (motion and electrodynamics) is richer than ordinary Minkovskian dynamics.  
\item Field equations (KG, Dirac,\ldots) in C-space.
\item A maximal force ( accelaration) principle and phase space duality are present in the theory. 
\item Is Maximal acceleration related to maximal complexity?
\item Born's reciprocal relativity +  Nambu dynamics and likely Finsler-like higher order geometries (sometimes referred as Kawaguchi geometry) seems to be relevant there.
\item A higher order M(inimal)-maximal $n$-order high derivative theory? We could formalate something like the $UR$ conjecture to guide us in the search of the theory of everything(if it exists).
\item No closed fundamental description of (super)string theory and M-theory from and invariance principle is avalaible yet. Almost 30 years ago, after the 2nd superstring theory revolution, dualities have challenged the way in which a theory is seen. 
\item Emergent spacetime needs likely some discrete grains. It seems no evidence of what could the fundamental degrees of freedom of spacetime be. They could be branes, strings, or some type of quantum preonic entities we don't know yet. But what are the symmetry (if any) group of these particles? Discrete-continuous group are important in some mathematical branches of mathematics (and they are key in moonshine conjecture or similar structures).

\end{enumerate}

Maybe, we need a radical reformulation of what (quantum) spacetime is. Even our notions of distance, time, energy and momentum could be changed and not easily described by real or complex numbers but other stranger class of numbers\cite{volovich}. Whatever the theory is behind QFT(QM) and GR as EFT, surely it has amazing mathematical structures yet to be discovered and found, and it certainly will change our actual preconceptions of what are space, time or fields, and it will be challenging and very hard to get hints of its origin and features. Physics is an experimental science, and we will require the full power of gravitational wave observatories, axion telescopes, and (if found) dark matter particle haloscopes to guess hints of the new fields, or, as well, to ask what is the new extended relativity principle of universal invariance, from which we could easily perform new verifications and tests of the nature of the dark matter and dark energy that arises, like the cosmological constant problem, and remains in the firs quarter of the 21st century. Experiments are leading now, after years of theoretical dominion, the searches for a more fundamental (invariant) description of Nature. Nevertheless, we should also push forward the findings of a symmetry principle enlarging and enhancing our current EFT, even when it is not an easy task.

\begin{center}
	\textbf{Acknowledgements}
\end{center}
These works (both the paper and the talk) are mainly based on the review \cite{CastroPavsicRev},  the paper \cite{Castro2010},  \textit{International Journal of Modern Physics of Modern Physics A}. Further references and details can be found in \cite{PavsicBook},  W.Pezzaglia [arXiv: gr-qc/9912025];I. R. Porteous, {\it  Clifford algebras and Classical Groups } (CUP, 1995);  S. Low: Jour. Phys { \textbf {A} } Math. Gen {\textbf {35}}, 5711 (2002);  
JMP. {\textbf {38}}, 2197 (1997); J. Phys.  {\textbf {A 40}}  (2007) 12095; ``Constraint Quantization of a worldline system invariant under
reciprocal Relativity, II'',  \url{https://arxiv.org/abs/0806.4794};  C. Castro, Phys Letts {\textbf {B 668}}  (2008) 442 and \textit{Notes on several phenomenological laws of quantum gravity} by Jean-Philippe Bruneton, ArXiv eprint, \url{https://arxiv.org/abs/1308.4044} . Also, I wish to thank C. C. Perelman for many discussions about ER anc C-spaces (he is a pioneer in that field), its Nambu-like geometry in the past years, and Matej Pavsic (he also pioneered the field of C-space as the arena for physics, reviewed in a book\cite{PavsicBook}) his kind invitation to the IARD 2016 workshop, where the original talk from which this paper arises was given.


\appendix 
\pagebreak 

\section{Hints of (beyond) ER in future past days}
Predictions on the rise of ER is not new. ER and BER are ideas that have been anticipated before. They visit us cyclically.\\
Notable recent examples:
\subsection{Preskill's hinting}

\begin{figure}[h!]
	\begin{center}
		\includegraphics[scale=0.5]{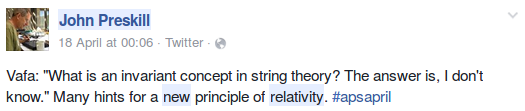}
	\end{center} 
	\caption{John Preskill's hinting a new relativity principle to come, also advanced by many others, e.g.,\cite{PavsicBook, Castro1, Castro2, Castro3,CastroFound}.}
\end{figure}

\subsection{IARD2012 Friedman's highlights}
Another one, from IARD, not long, long ago\cite{YaakovFriedman}:

\begin{figure}[h!]
	\begin{center}
\includegraphics[scale=0.25]{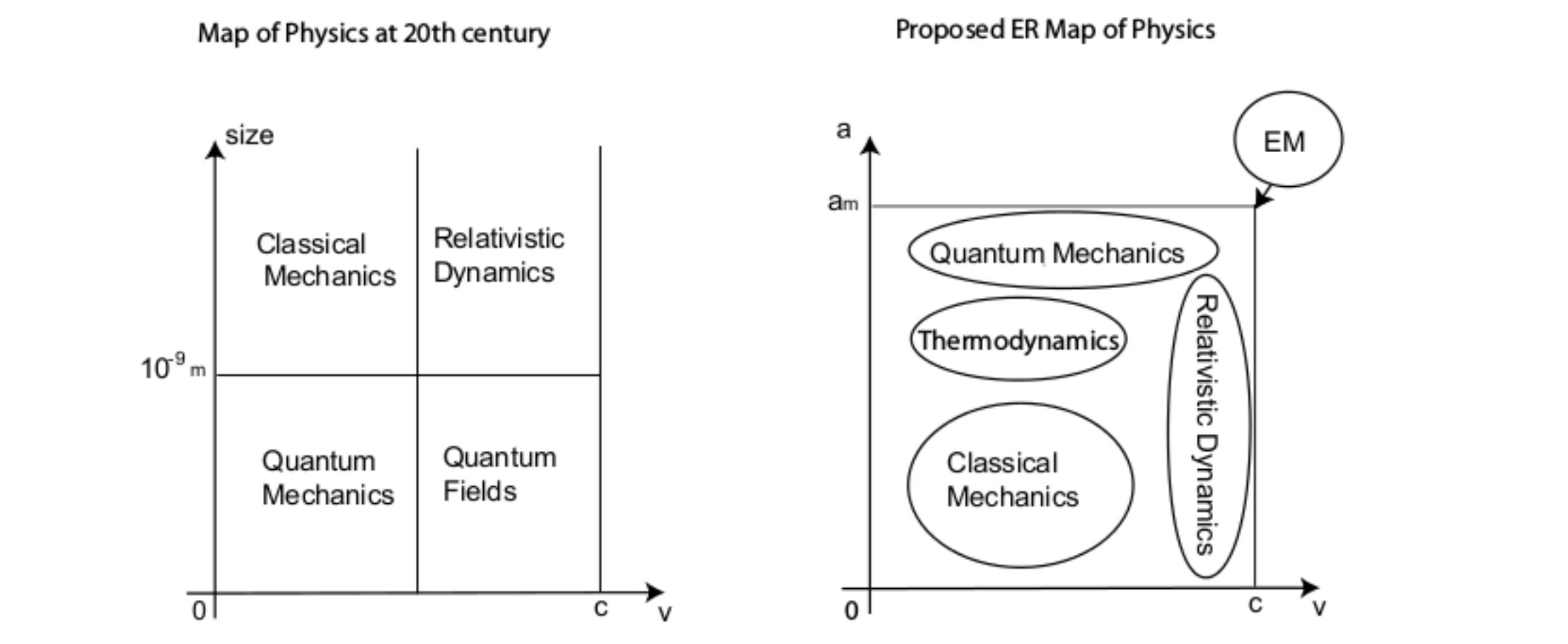}

\end{center} 
\caption{The map of Extended Relativity from Y. Friedman in \cite{YaakovFriedman}.}
\end{figure}

\pagebreak
\subsection{Caianiello's \& Brandt maximal acceleration relativity}

Caianiello \cite{Caianiello1981,Caianiello1982} and Brandt \cite{Brandt1983} pioneered studies on the maximal acceleration principle
\begin{figure}[h!]
	\begin{center}
		\includegraphics[scale=0.5]{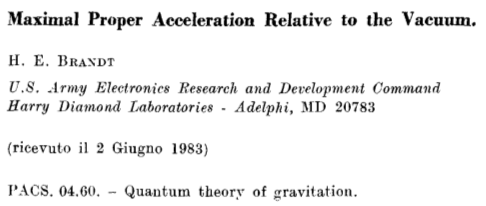}\\
		\includegraphics[scale=0.5]{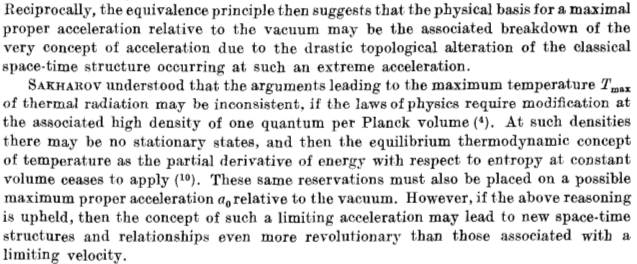} 
	\end{center}
	\textbf{Note the coolest and most important remark from above:}\\
	\begin{center}
		\includegraphics[scale=0.75]{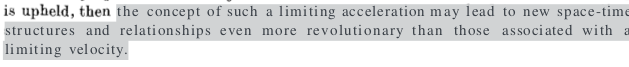}
	\end{center} 
	\caption{Caianiello and Brandt anticipating as well a new relativity principle.}
\end{figure}

\pagebreak

\section{Ultimate relativity: a mind map}

\begin{figure}[h!]
	\begin{center}
		\includegraphics[scale=0.3]{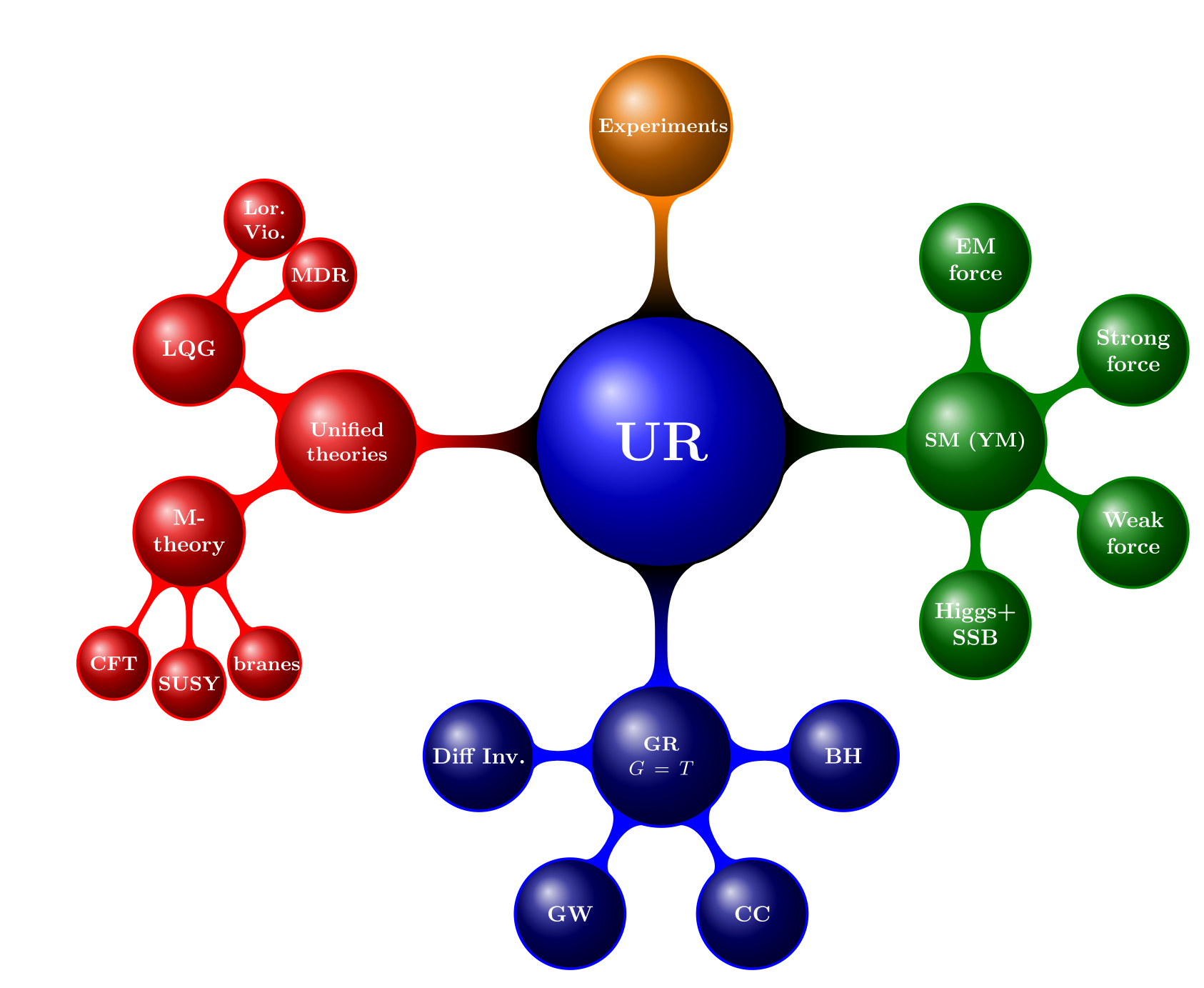}
	\end{center}
	
	\caption{UR, a mind map.}
\end{figure}

\pagebreak
\section{Beyond ER: a higher order M(in-Max)-theory of ER?}

\textbf{Question:} What about a generalized relativistic dynamics 
 for $E=\Gamma mc^2$, using ``duality'' and ``symmetry'', such any derivative appears on equal footing? Say
\[\Gamma (X^2,V^2,A^2,\ldots)=\dfrac{\sqrt{1-\dfrac{l_0^2}{X^2}}\sqrt{1-\dfrac{c_0^2}{V^2}}\sqrt{1-\dfrac{a_0^2}{A^2}}\cdots}{\sqrt{1-\dfrac{X^2}{L_\Lambda^2}}\sqrt{1-\dfrac{V^2}{C^2}}\sqrt{1-\dfrac{A^2}{A^2_m}}\cdots }\]
Can we test it? Is it crazy enough to be true or useful for high derivative theories? \textit{Note that} Caianiello's epitachyons are entities with $A>A_m$. Maybe too many constants? Not at all: likely, maximal and minimal acceleration, jerk,\ldots are derived from fundamental maximal and minimal lengths $l_0,L_\Lambda$. Who knows?\\

\end{document}